\documentclass[%
 reprint,
 amsmath,amssymb,
 aps,
]{revtex4-1}

\usepackage{float}
\usepackage{graphicx}% Include figure files
\usepackage{dcolumn}% Align table columns on decimal point
\usepackage{bm}% bold math
\usepackage[T1]{fontenc}
\usepackage{textcomp}
\usepackage[outercaption]{sidecap}
\begin{document}

\title{Mechanically-induced disorder in CaFe$_2$As$_2$: a $^{57}$Fe M\"ossbauer study}

\author{Xiaoming Ma$^{1,2}$, Sheng Ran$^{1}$, Paul C. Canfield$^{1}$, and Sergey L. Bud'ko$^{1}$}
\email{Corresponding author: budko@ameslab.gov}

\affiliation{ $^1$Ames laboratory, US DOE, and Department of Physics and Astronomy, Iowa State University, Ames, Iowa 50011, USA.\\$^2$Institute of Applied Magnetics, Key Laboratory for Magnetism and Magnetic Materials of the Ministry of Education, Lanzhou University, Lanzhou, Gansu Province, 730000, China.}

\date{\today}

\begin{abstract}
$^{57}$Fe M\"ossbauer spectroscopy was used to perform a microscopic study on the extremely pressure and strain sensitive compound, CaFe$_2$As$_2$, with different degrees of strain introduced by grinding and annealing. At the base temperature, in the antiferromagnetic/orthorhombic phase, compared to a sharp sextet M\"ossbauer spectrum of single crystal CaFe$_2$As$_2$, which is taken as an un-strained sample, an obviously broadened sextet and an extra doublet were observed for ground CaFe$_2$As$_2$ powders with different degrees of strain. The M\"ossbauer results suggest that the magnetic phase transition of CaFe$_2$As$_2$ can be inhomogeneously suppressed by the grinding induced strain to such an extent that the antiferromagnetic order in parts of the grains forming the  powdered sample remain absent all the way down to 4.6 K. However, strain has almost no effect on the temperature dependent hyperfine magnetic field in the grains with magnetic order. Additional electronic and asymmetry information was obtained from the isomer shift and quadrupole splitting. Similar isomer shift values in the magnetic phase for samples with different degrees of strain, indicate that the stain does not bring any significant variation of the electronic density at $^{57}$Fe nucleus position. The absolute values of quadrupole shift in the magnetic phase decrease and approach zero with increasing degrees of strain, indicating that the strain reduces the average lattice asymmetry at Fe atom position.

\begin{description}
\item[PACS numbers]74.70.Xa, 76.80.+y, 74.62.Dh
\end{description}
\end{abstract}


\maketitle

%\tableofcontents

\section{\label{sec:level1}Introduction}

As one of the parent compounds for iron based superconductors, CaFe$_2$As$_2$ \cite{NiNiFirst}, manifests an extreme example of strongly coupled, first order, structural and magnetic transitions\cite{NiNiFirst,Coupling}. At $\sim 170$ K CaFe$_2$As$_2$ undergoes transition from a high temperature paramagnetic tetragonal phase to a low temperature antiferromagnetic orthorhombic phase. Furthermore, it is one of the most pressure sensitive Fe-As based materials: it takes less than 0.5 GPa at low temperatures to induce a transition from an antiferromagnetic, tetragonal state to paramagnetic, collapsed tetragonal phase \cite{Milton1, Yu, scattering}. If the pressure medium is hydrostatic, no superconductivity is observed under pressure, but when some non-hydrostatic component is present (as in piston-cylinder cells with  liquid media at room temperature and solid at low temperatures) superconductivity, most likely filamentary, can be induced \cite{Milton1, Yu,  Milton2, LANL, scattering, PaulPhysicaC}. The pressure/stress response of several phase transitions in Co-substituted CaFe$_2$As$_2$ \cite{Sheng2}  is very anisotropic \cite{TE} with hydrostatic pressure derivatives reaching  record high values for an inorganic compound \cite{gati}.

The physical properties and the ground state of CaFe$_2$As$_2$ single crystals are also remarkably dependent on the growth procedure \cite{Sheng1,ShengThesis}. The annealing/quenching temperature can affect the strain in the crystals and further tune the structure and magnetic properties of CaFe$_2$As$_2$. The crystals grown out of FeAs flux, quenched from high temperature, exhibit a transition from the paramagnetic, tetragonal phase to the nonmagnetic, collapsed tetragonal phase, in contrast to the behavior of CaFe$_2$As$_2$ grown from Sn flux  \cite{Sheng1,ShengThesis}. The process of annealing and quenching can be used as an additional control parameter which can tune the ground state of CaFe$_2$As$_2$ systematically. The detailed phase diagram of transition temperature, versus annealing/quenching temperature determined by electrical resistance and magnetic susceptibility measurements can be found in Refs.\cite{Sheng1, ShengThesis}. The effects of annealing and quenching are similar to the effects of applied pressure \cite{gati,Sheng1,ShengThesis}.

Transition metal substitution for Fe combined with judicious annealing/quenching result allow to tune CaFe$_2$As$_2$ to have antiferromagnetic orthorhombic, superconducting tetragonal, paramagnetic tetragonal, and paramagnetic collapsed tetragonal ground states, showing a great sensitivity of this material to perturbations. Even different sample mounting procedures may cause significant shift in transition temperatures \cite{ShengThesis}.

\begin{figure}[htp]
\includegraphics[width=0.5\textwidth]{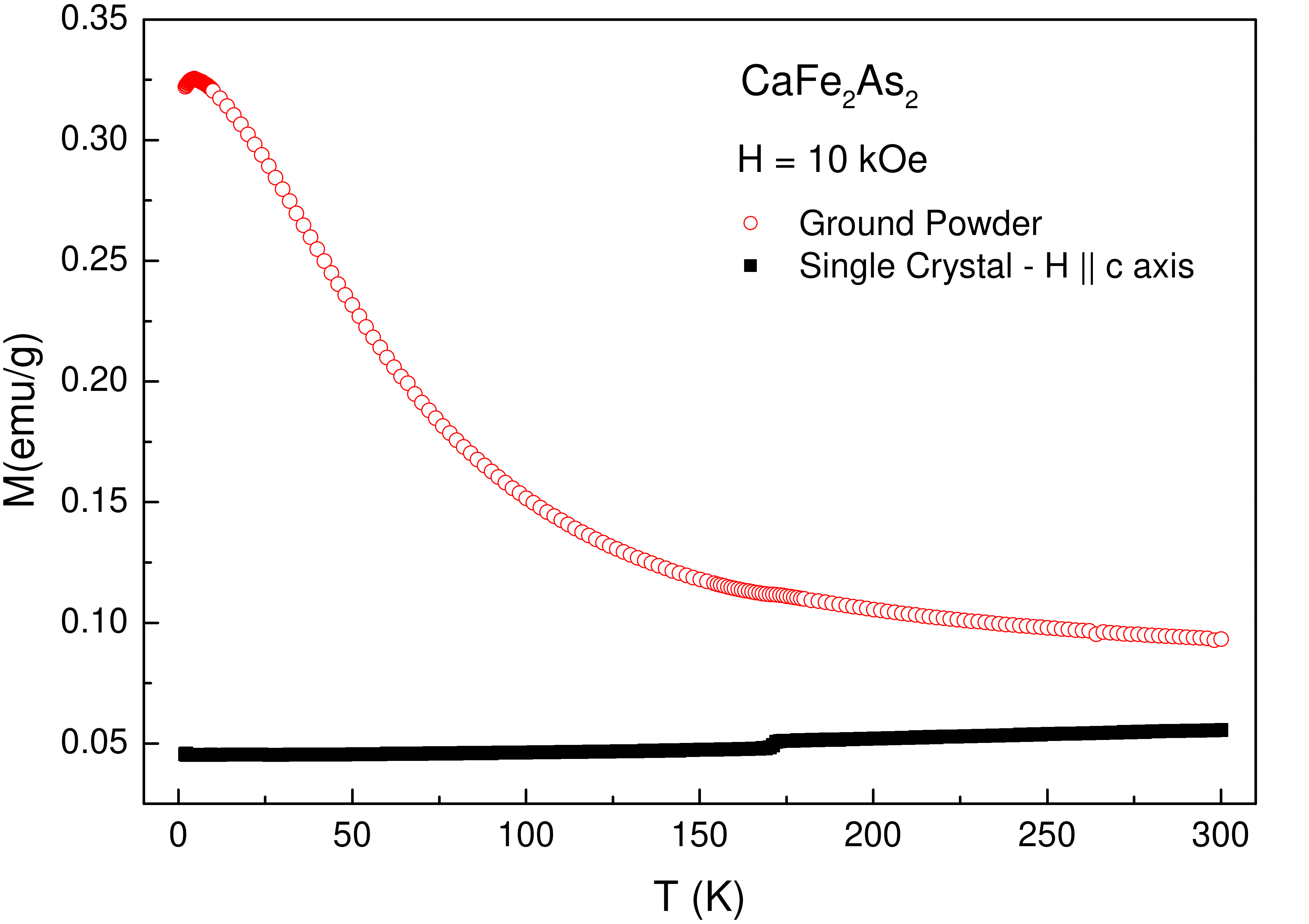}
\caption{\label{fig:epsart} (Color online) Temperature dependent magnetization  of CaFe$_2$As$_2$ single crystal with applied magnetic field along $c$ axis and ground CaFe$_2$As$_2$ powder.}
\end{figure}

\begin{figure}[htp]
\includegraphics[width=0.5\textwidth]{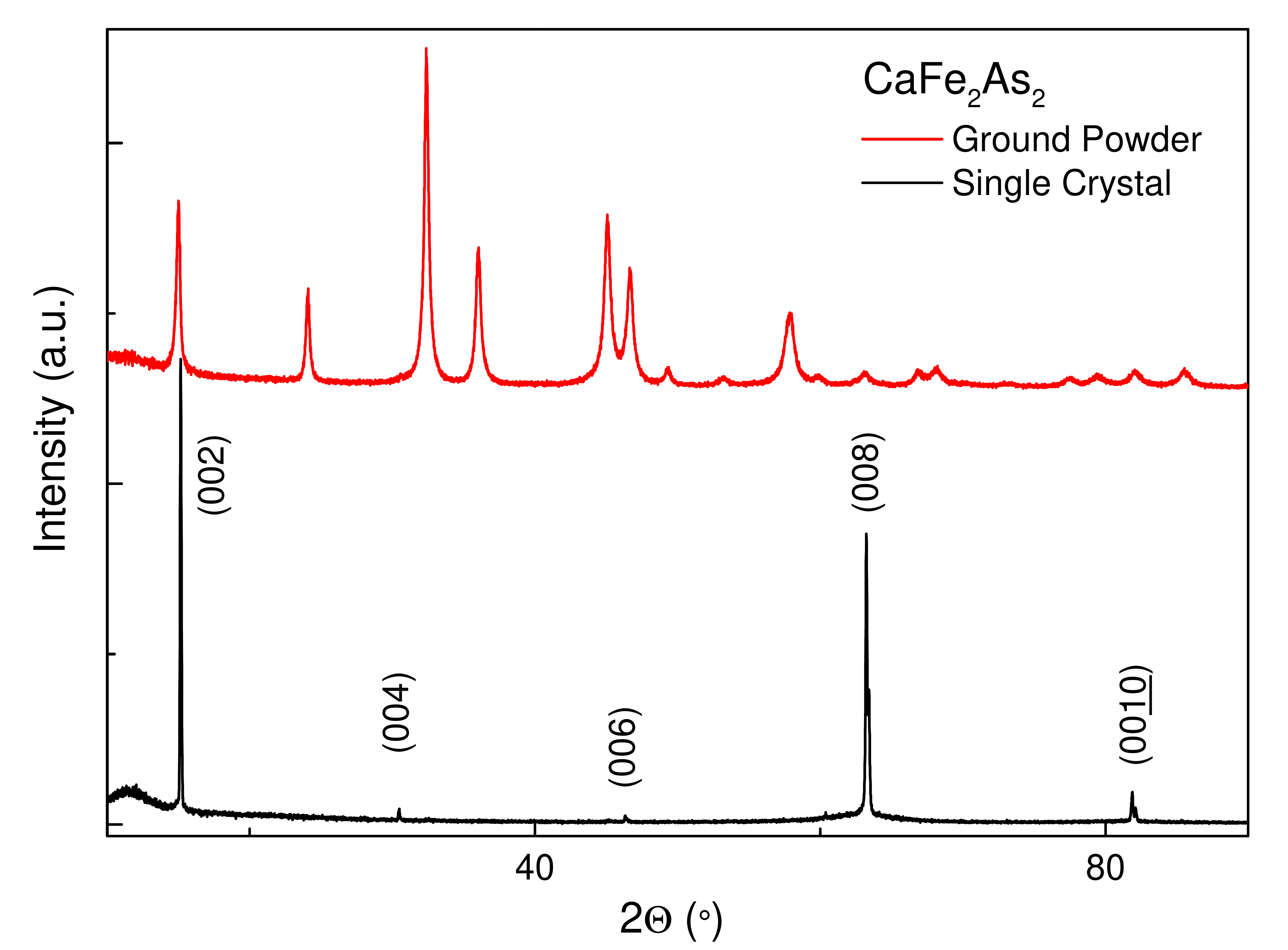}
\caption{\label{fig:epsart} (Color online) XRD results of CaFe$_2$As$_2$ single crystal plate and ground powder.}
\end{figure}

CaFe$_2$As$_2$ is so sensitive to pressure that grounding of single crystals  by hand in a mortar can cause significant variation of magnetic and structural properties of this compound. Fig. 1 presents a comparison of the temperature dependent magnetization of CaFe$_2$As$_2$ single crystal plate and ground powder. The magnetization of the ground powder is considerably higher than that of the single crystal over the whole observed temperature range, and a Curie tail can be clearly observed. The X-ray diffraction (XRD) spectra of the single crystal CaFe$_2$As$_2$ and the ground powder are shown in Fig.2. The diffraction peaks of the ground powder are significantly broader. Bulk, macroscopic measurements can give only limited information on effects of mechanically induced disorder on  CaFe$_2$As$_2$. Therefore, we employ a local probe, $^{57}$Fe  M\"ossbauer spectroscopy, to study how the microscopic structural and magnetic properties are affected by the strain induced by mechanical grinding.

In this work, we present a systematic $^{57}$Fe  M\"ossbauer study of CaFe$_2$As$_2$ samples with different degrees of strain: single crystals, as ground powder samples and annealed powders. Evolution of the $^{57}$Fe M\"ossbauer spectra in the temperature range 4.6 K to 300 K and corresponding hyperfine parameters will be discussed.

\section{Experimental details}
Single crystals of CaFe$_2$As$_2$ were grown out from an Sn flux, using conventional high-temperature solution method \cite{CanfieldFisk}.  Elements Ca, Fe, As and Sn were combined together in the ratio of Ca : Fe : As : Sn = 2 : 3.5 : 4 : 96. Single crystals were grown by slowly cooling the Ca-Fe-As-Sn melt from $1180\,^{\circ}{\rm C}$ to $600\,^{\circ}{\rm C}$ at $5\,^{\circ}{\rm C}$/h, and then decanting off the excess liquid flux. Detailed description of the crystal growth can be found elsewhere \cite{NiNiFirst}. The obtained crystals have typical size of 5$\times$5$\times$1 mm$^3$ and will be referred to as "Sngl. Crys.". Some CaFe$_2$As$_2$ single crystals were ground into fine powder in an agate mortar by hand at room temperature (RT) and the obtained CaFe$_2$As$_2$  powder will be referred to as "Grnd. Pwd.*"  In order to investigate the effect of annealing, used to release some strain in the powdered sample, first we ground some CaFe$_2$As$_2$ single crystals into powder in a glove box, so as to avoid  oxidation, then part of the powder was sealed in a fused silica tube under the protection of an argon atmosphere, and annealed at $400\,^{\circ}{\rm C}$ for 3 days. This batch of ground powder and annealed ground powder are going to be referred to as "Grnd. Pwd." and "Ann. Pwd.", respectively.

XRD measurements of the plate-like single crystal and powder samples were performed at RT using a Rigaku Miniflex diffractometer with Cu K$\alpha$ radiation. The XRD spectrum of the single crystal plate is shown in Fig. 2. As can be seen, only (00$l$) diffraction peaks can be observed, which indicates that the crystallographic $c$-axis is perpendicular to the plane of the plate. The good quality of the CaFe$_2$As$_2$ single crystal was confirmed by temperature-dependent magnetization and resistivity measurements. Clear features, including a sharp drop in magnetization and a sharp jump in resistivity, take place upon cooling through the transition temperature, 171 K, which is associated with the structural and magnetic phase transition and is consistent with the previous reported results \cite{NiNiFirst,Sheng1}. To estimate the lattice parameters and the amount of secondary phase, the powder X-ray spectra were refined by Rietveld analysis using the EXPGUI software \cite{EXPGUI}.

$^{57}$Fe M\"ossbauer spectroscopy measurements were performed using a SEE Co. conventional constant acceleration type spectrometer in transmission geometry with an $^{57}$Co(Rh) source, that had an initial intensity 50 mCi (2 years ago) and was kept at RT. The spectrometer was calibrated with an $\alpha$-Fe foil at RT. For the CaFe$_2$As$_2$ single crystals, the absorber was prepared by arranging plate-like single crystals with dimensions of $\sim$2$\times$2$\times$(0.05-0.1) mm$^3$ to form a mosaic disc with an area of $\sim 1.5$ cm$^2$. The absorbers of powder samples were made by mixing the compounds and vacuum grease homogeneously and then squeezing them between two pieces of papers to form a uniform disc, which contained $\sim$5 mg of natural Fe/cm$^2$. The cooling of the absorbers were achieved using Janis model SHI-850-5 closed cycle refrigerator with vibration damping. The isomer shift (IS) is measured relative to the $\alpha$-Fe foil at RT.

\section{Results and discussion}

Fig. 3 presents the magnetization versus temperature results of Sngl. Crys., Grnd. Pwd.*, Grnd. Pwd. and Ann. Pwd. in an applied 10 kOe magnetic field. Their respective temperature derivatives near transition temperature are shown in Fig. 4. Considering different preparation of the samples, it is reasonable to speculate that Sngl. Crys. has a minimum strain, Grnd. Pwd.* and Grnd. Pwd. have a maximum strain and the strain in the Ann. Pwd. is between Sngl. Crys. and Grnd. Pwd. In the Fig. 3, the magnetization values of all the three powder sample are much higher than the single crystal sample over the whole temperature range so it appears that the higher the strain is, the higher the magnetization value is.  The strain also has impact on the behavior of the structural-magnetic phase transition. As shown in Fig. 4, quite different from the sharp peak of the Sngl. Crys. sample, the magnetization result of Grnd. Pwd. and Grnd. Pwd.* barely show the structural-magnetic phase transition feature around 170 K. For the Ann. Pwd., because some strain is released after the annealing procedure, the feature associated with the phase transition emerges  again, at slightly higher temperature of 175 K.

\begin{figure}[htp]
\centering
\includegraphics[width=0.5\textwidth]{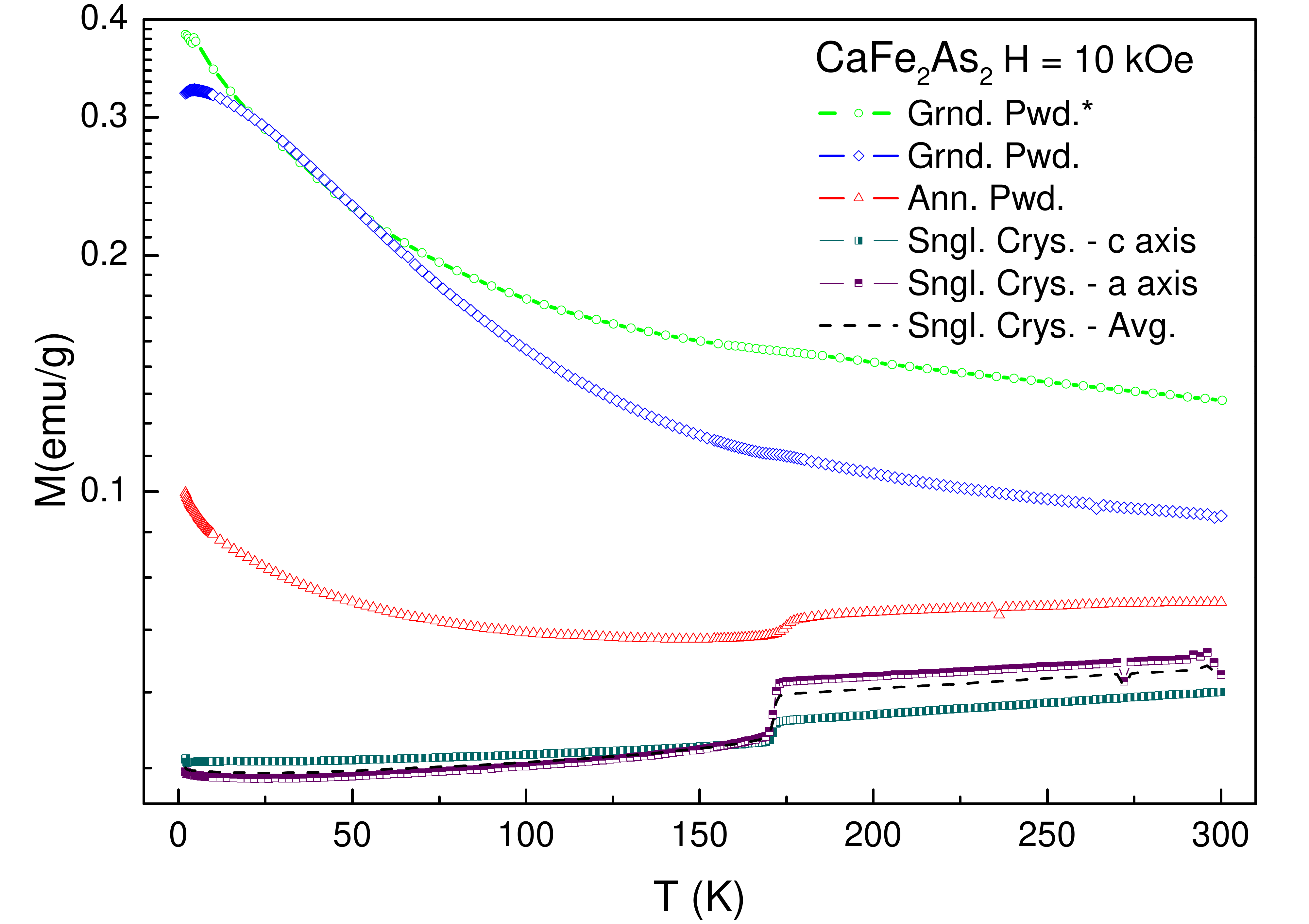}
\caption{\label{fig:epsart} (Color online) Temperature dependent magnetization results of Grnd. Pwd.*, Grnd. Pwd. Ann. Pwd. and Sngl. Crys. The detailed information about the samples is presented in the text. The measurements were performed at an external 10 kOe magnetic field. For the Sngl. Crys. sample, measurements were performed with the field applied along $a$-axis and $c$-axis separately and the black dash line represents the polycrystalline average result of the two directions.}
\end{figure}

\begin{figure}[htp]
\centering
\includegraphics[width=0.5\textwidth]{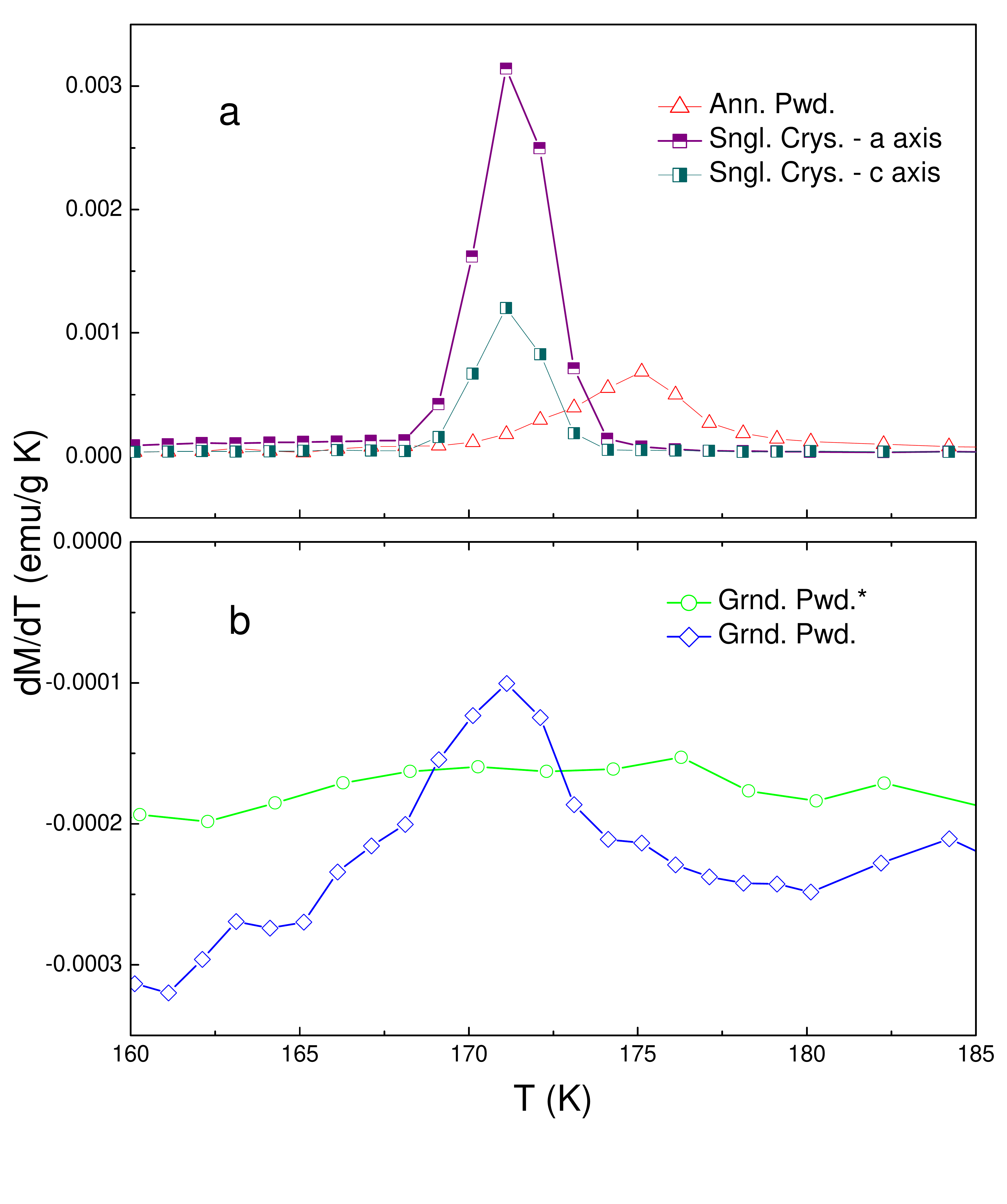}
\caption{\label{fig:epsart} (Color online) Derivative of magnetization, $dM/dT$, of (a) Sngl. Crys. and Ann. Pwd. and (b) Grnd. Pwd.* and Grnd. Pwd. in the transition temperature region.}
\end{figure}

We performed XRD measurements on all the samples. The results of powder CaFe$_2$As$_2$ samples, Grnd. Pwd.*, Grnd. Pwd. and Ann. Pwd. are grossly similar to the pattern shown in Fig.2. The  lattice parameters of powder samples obtained by Rietveld refinement are listed in Table 1.  

\begin{table}[ht]
\caption[]{Lattice parameters of CaFe$_2$As$_2$ samples, Grnd. Pwd.*, Grnd. Pwd. and Ann. Pwd., obtained through the Rietveld refinement of XRD results of RT.}
\begin{ruledtabular}
\begin{tabular}{lcccc}
Sample          &a             &c         &z        &a/c    \\
               &\mbox{\AA}   &\mbox{\AA} &          &       \\
\hline
Grnd. Pwd.*    &3.8908(4)     &11.775(1) &0.3658(1) &0.3304 \\
Grnd. Pwd.     &3.8980(7)     &11.799(2) &0.3674(9) &0.3304 \\
Ann. Pwd.      &3.8894(3)     &11.797(1) &0.36715(8)&0.3297 \\
\end{tabular}
\end{ruledtabular}
\end{table}

\begin{figure}[htp]
\includegraphics[width=0.5\textwidth]{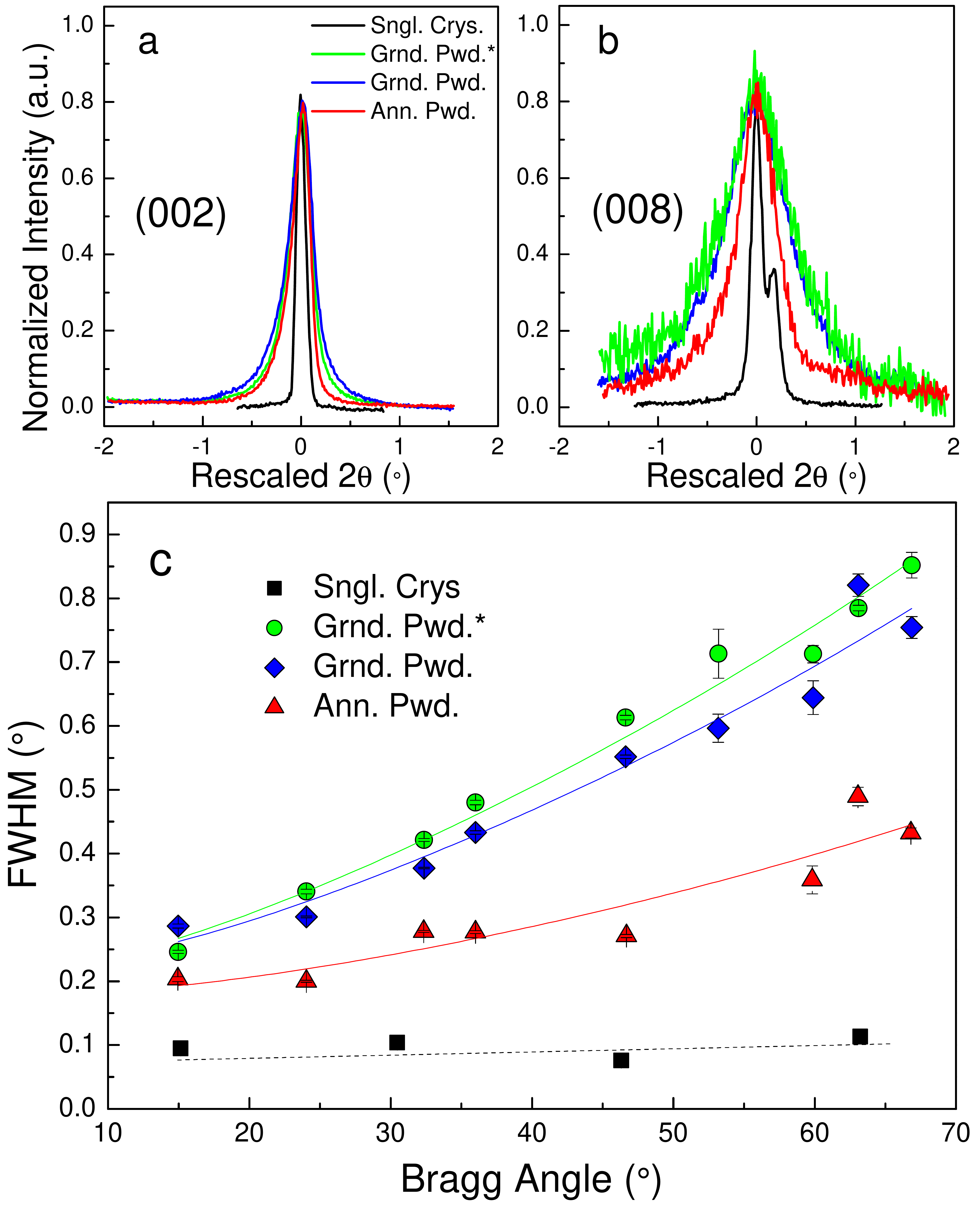}
\caption{\label{fig:epsart} (Color online) The top panels, (a) and (b), show  normalized diffraction peaks of (002) and (008) for Sngl. Crys., Grnd. Pwd.*, Grnd. Pwd. and Ann. Pwd., respectively (the $x$ axis is rescaled by subtracting the Bragg angle of each peak); (c) plots of the peak widths (FWHM) at various diffraction angles. The least squares fit results to Eq.(1) are shown by solid lines. The black dashed line through Sngl.Crys. data is a guide to the eye.}
\end{figure}

\begin{figure*}[htp]
\includegraphics[width=1\textwidth]{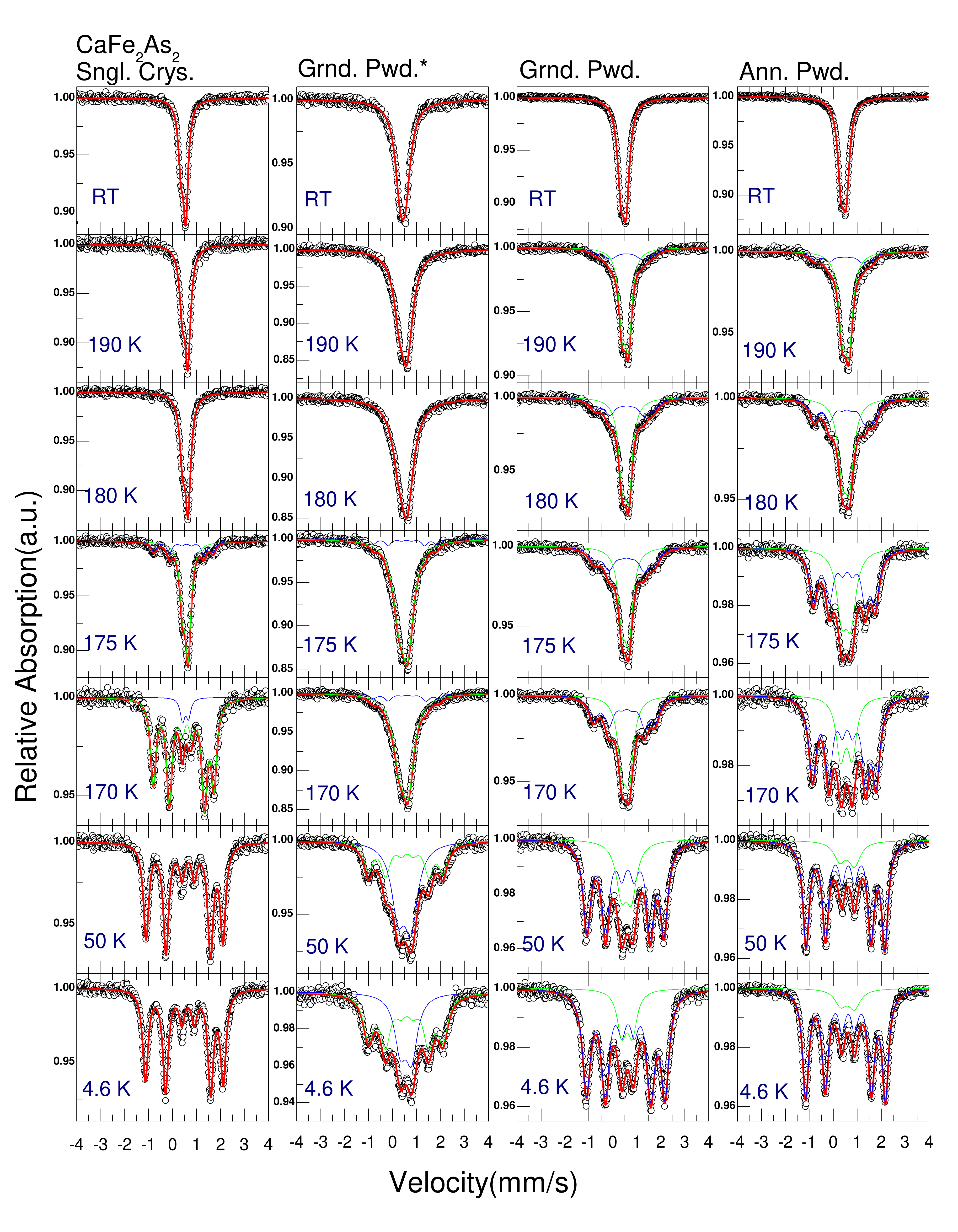}
\caption{\label{fig:epsart} (Color online) $^{57}$Fe M\"ossbauer spectra symbols) and fits (lines) of Sngl. Crys., Grnd. Pwd.*, Grnd. Pwd. and Ann. Pwd. M\"ossbauer spectra at different temperatures.}
\end{figure*}

\begin{table*}
\centering
\caption[]{The summary of hyperfine parameters of the four investigated CaFe$_2$As$_2$ samples at 4.6 K and RT.}
\begin{ruledtabular}
\begin{tabular}{lccccccc}
Sample         &T    &        &IS         &QS         &HF       &Linewidth   &Rel. Area\\
               &\mbox{K}&     &\mbox{mm/s}&\mbox{mm/s}&\mbox{T} &\mbox{mm/s} &\mbox{\%}  \\
\hline
Sngl. Crys.    &RT   &doublet &0.448(2)   & 0.202(2)  &---      &0.254(3)    &100      \\
               &4.6  &sextet  &0.5708(7)  & -0.159(1) &10.078(6)&0.280(2)    &100      \\
Grnd. Pwd.*    &RT   &doublet &0.429(2)   &0.223(7)   &---      &0.520(9)    &100      \\
               &4.6  &doublet &0.5557(7)  &0.467(8)   &---      &0.58(2)     &37.5    \\
               &     &sextet  &0.574(4)   &-0.055(6)  &9.64(3)  &0.51(1)     &62.5      \\
Grnd. Pwd.     &RT   &doublet &0.443(1)   &0.219(1)   &---      &0.341(2)    &100     \\
               &4.6  &doublet &0.593(5)   &0.441(7)   &---      &0.46(2)     &17.5     \\
               &     &sextet  &0.586(1)   &-0.097(2)  &10.157(9)&0.938(3)    &82.5     \\
Ann. Pwd.      &RT   &doublet &0.445(1)   &0.216(1)   &---      &0.353(2)    &100     \\
               &4.6  &doublet &0.60(1)    &0.57(2)    &---      &0.60(5)     &12.0     \\
               &     &sextet  &0.5738(6)  &-0.117(1)  &10.317(5)&0.317(2)    &88.0     \\
\end{tabular}
\end{ruledtabular}
\end{table*}

Compared with the single crystal data,  diffraction peaks broadening can be observed for powder samples as shown for the normalized  (002) and (008) diffraction peaks  in Fig. 5 (a) and (b). As expected, the high-angle (008) diffraction peak shows more obvious broadening than the lower angle (002) diffraction peak. We can analyze the peak width as follows. The values of the full width at half maximum of distinguishable diffraction peaks for each sample versus Bragg angle (2$\theta$) are summarized in Fig. 5 (c). Usually, line broadening comes from the instrument parameters, the finite crystallite size and the lattice strain. For conventional diffractometer, when the crystallite size is under 100 nm, the line broadening caused by crystallite size effect can be detected and well described by Scherrer formula \cite{Scherrer}. Considering (a) the almost angle-independent linewidth in Sngl. Crys. in Fig. 5 and (b) the typical micron size material obtained by grinding, we believe that the broadening mainly comes from a contribution due to the lattice strain. Strain is defined as the deformation of an object divided by it's ideal length, $\Delta$d/d. In crystals there are two types of strain: uniform strain and non-uniform strain. Uniform strain simply leads to a change in the unit cell parameters and shift of the peaks and doesn't contribute to broadening. Non-uniform strain leads to systematic shifts of atoms from their ideal positions and to peak broadening, which was first pointed out by Stokes and Wilson \cite{Stokes}. In order to extract information about the microstructure of materials and support our deduction, we analyzed the data from our powder samples by a combination of Scherrer formula and the model given by Stokes and Wilson \cite{equation1}:

\begin{equation}
\beta^2 = [\frac{0.9\lambda}{Dcos\theta}]^2 + [4\varepsilon tan\theta]^2 + \beta^2_0,
\end{equation}

\noindent where $\beta$ is the total broadening, $\lambda$ is the wavelength of the X-ray, D is the crystallite size, $\beta_0$ is the instrumental broadening and $\varepsilon$ is the strain.

The  first term is very close to zero and $\beta_0$ is approximately 0.03$^{\circ}$. This means that there is almost no contribution from the size effect and, compared with the observed line broadening, the instrument broadening can also be neglected. The calculated $\varepsilon$ is 0.0055(1), 0.0050(2) and 0.0027(4) for Grnd. Pwd.*, Grnd. Pwd. and Ann. Pwd., respectively. The grinding outside of the glove box was more aggressive, so the strain value in Grnd. Pwd.* is slightly larger than that in Grnd. Pwd. After the heat treatment, nearly half of the non-uniform strain was eliminated.

The microscopic properties of Sngl. Crys., Grnd. Pwd.*, Grnd. Pwd. and Ann. Pwd. were studied by $^{57}$Fe M\"ossbauer spectroscopy in the temperature range of 4.6 K to 300 K, some representative spectra are shown in Fig 6.
Similar to previous reports \cite{CaMoss1,CaMoss2}, the M\"ossbauer spectrum of  single crystalline CaFe$_2$As$_2$ at RT has two sharp asymmetric  lines, which can be well fitted by a doublet with two different intensity peaks. The asymmetry comes from the fact that the relative peak areas of M\"ossbauer quadrupole doublets are a function of the angle between the $\gamma$ ray and the electric field gradient (EFG). The M\"ossbauer spectra of the powder samples Grnd. Pwd.*, Grnd. Pwd. and Ann. Pwd. at RT present doublets with a little asymmetry. The slight asymmetry indicates the existence of preferential orientation in the powder samples. Therefore we employed a doublet with free intensity ratio parameter of the two peaks to fit the paramagnetic phase and a sextet with free intensity ratio of the 2(5) peak to the 3(4) peak for the magnetic phase (with peaks numbered 1 - 6 from lowest to highest velocity in Fig. 6). Several spectra of Grnd. Pwd. and Ann. Pwd. near the transition temperature were fitted with the constraint that the intensity ratio of 2(5) and 3(4) peaks is smaller than 4. The fitted results are presented by solid lines in Fig. 6 as well. The corresponding hyperfine parameters at 4.6 K and RT of all the four samples are summarized in Table 2. It should be noted that the "Mixed M + Q Static Hamiltonian" model is physically  more appropriate for fits of the magnetic spectra, however, since for the powder samples (a) two subspectra, sextet and doublet, are needed to fit the data, and (b) preferential orientation is present in the powder absorbers (see below), we have chosen an approach with less fitting parameters that we believe still captures the essence of the data.

\begin{figure}[htp]
\includegraphics[width=0.5\textwidth]{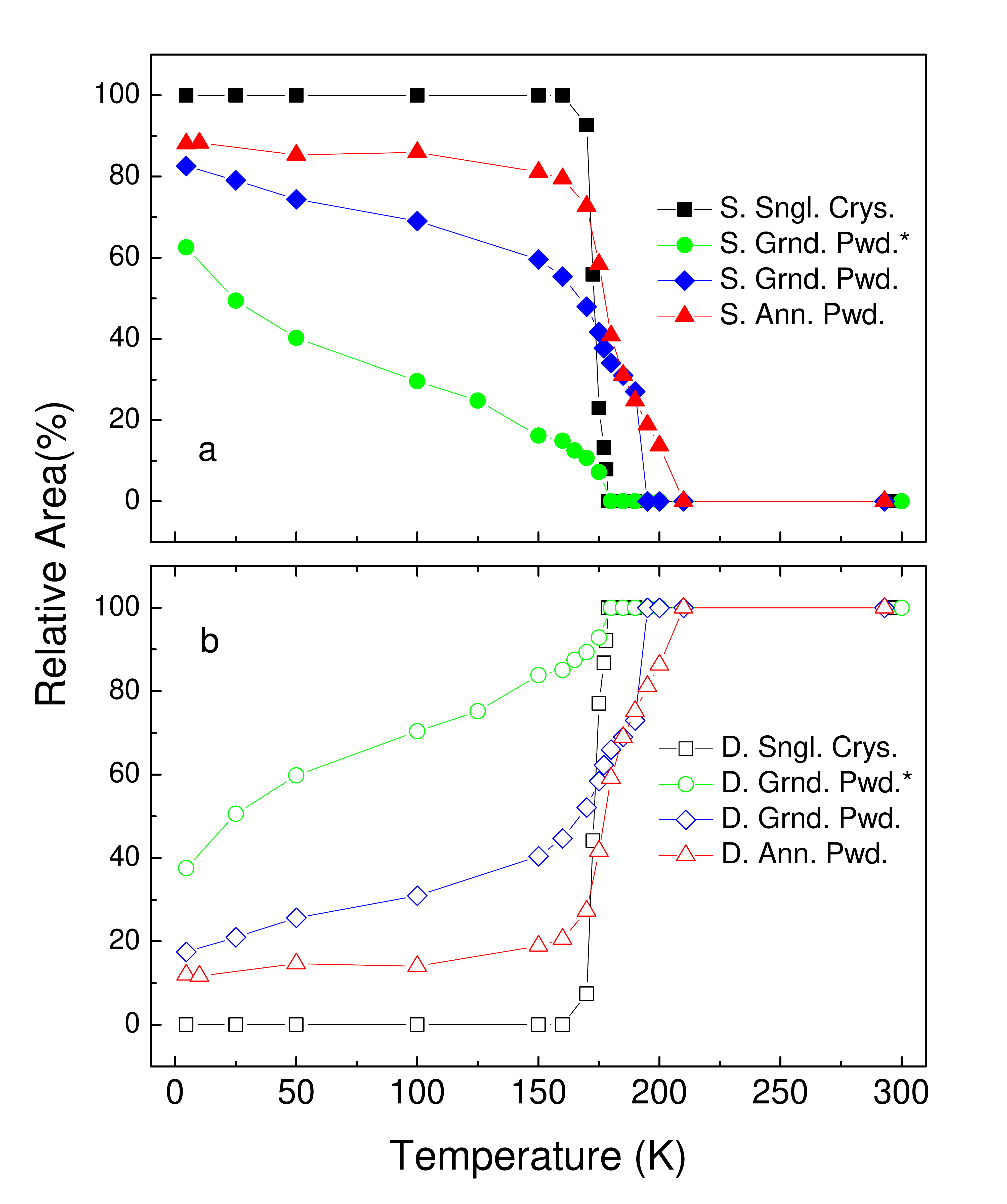}
\caption{\label{fig:epsart} (Color online) Temperature dependent relative area of the (a) sextet and (b) doublet sub-spectra of all the investigated samples.}
\end{figure}

\begin{figure}[htp]
\includegraphics[width=0.5\textwidth]{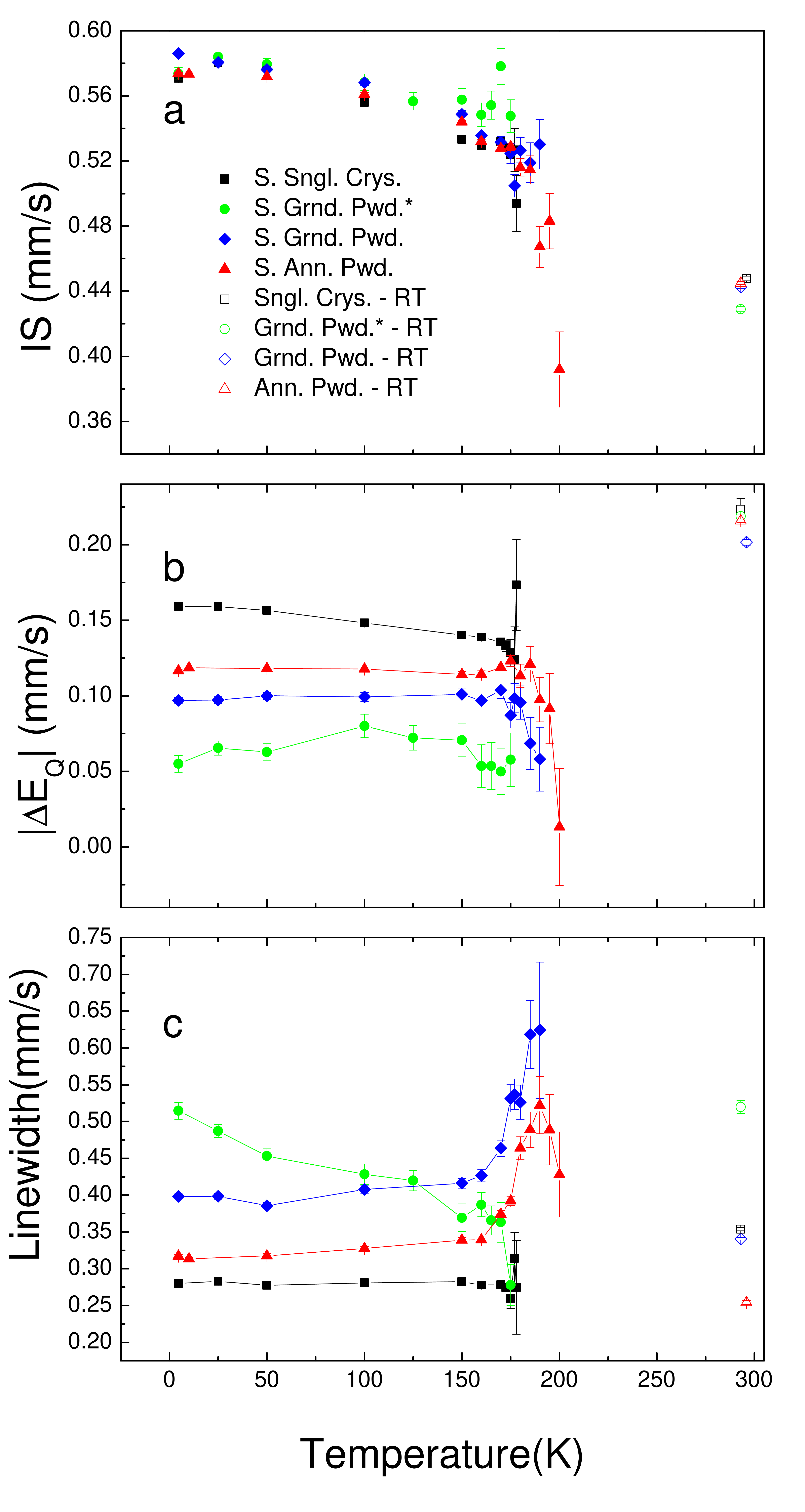}
\caption{\label{fig:epsart} (Color online) Temperature dependent hyperfine parameters, (a) isomer shift, (b) quadrupole splitting and (c) linewidth, of the sextet of all the four investigated samples.}
\end{figure}

As shown in Fig. 6, one significant effect of grinding is the presence of a doublet accompanying the sextet in all the three ground samples at temperatures much lower than the structural/magnetic transition temperature. We summarized the variation of the relative area of the sextet and the doublet with temperature in Fig.7 (a) and (b), respectively. For Sngl. Crys., the area of the doublet decreases sharply at 175 K and goes to 0 \% at 160 K. In contrast, for the samples that experienced grinding: Grnd. Pwd.*, Grnd. Pwd. and Ann. Pwd., the areas of the doublets all decrease at a slower rate and doublets are still present in the spectra even at 4.6 K. The amount of the paramagnetic phase has relationship with the degree of strain. At 4.6 K, the relative area of the doublet is 12 \%, 17.5 \% and 37.5 \% for Ann. Powd., Grnd. Pwd. and Grnd Pwd.*, respectively. The higher the strain in the compound is, the larger the area of the doublet is. It is reasonable to attribute the sextet sub-spectrum to the grains of the sample with mild strain and the doublet sub-spectrum to the grains with significant strain/disorder. These seriously disordered/damaged grains are most likely also responsible for the high background of the magnetization in Fig. 3. In addition, the near linear decrease of the relative area of the doublet also suggests a distribution of the phase transition temperatures caused by the distribution of the degree of strain in those samples.

\begin{figure}[htp]
\includegraphics[width=0.5\textwidth]{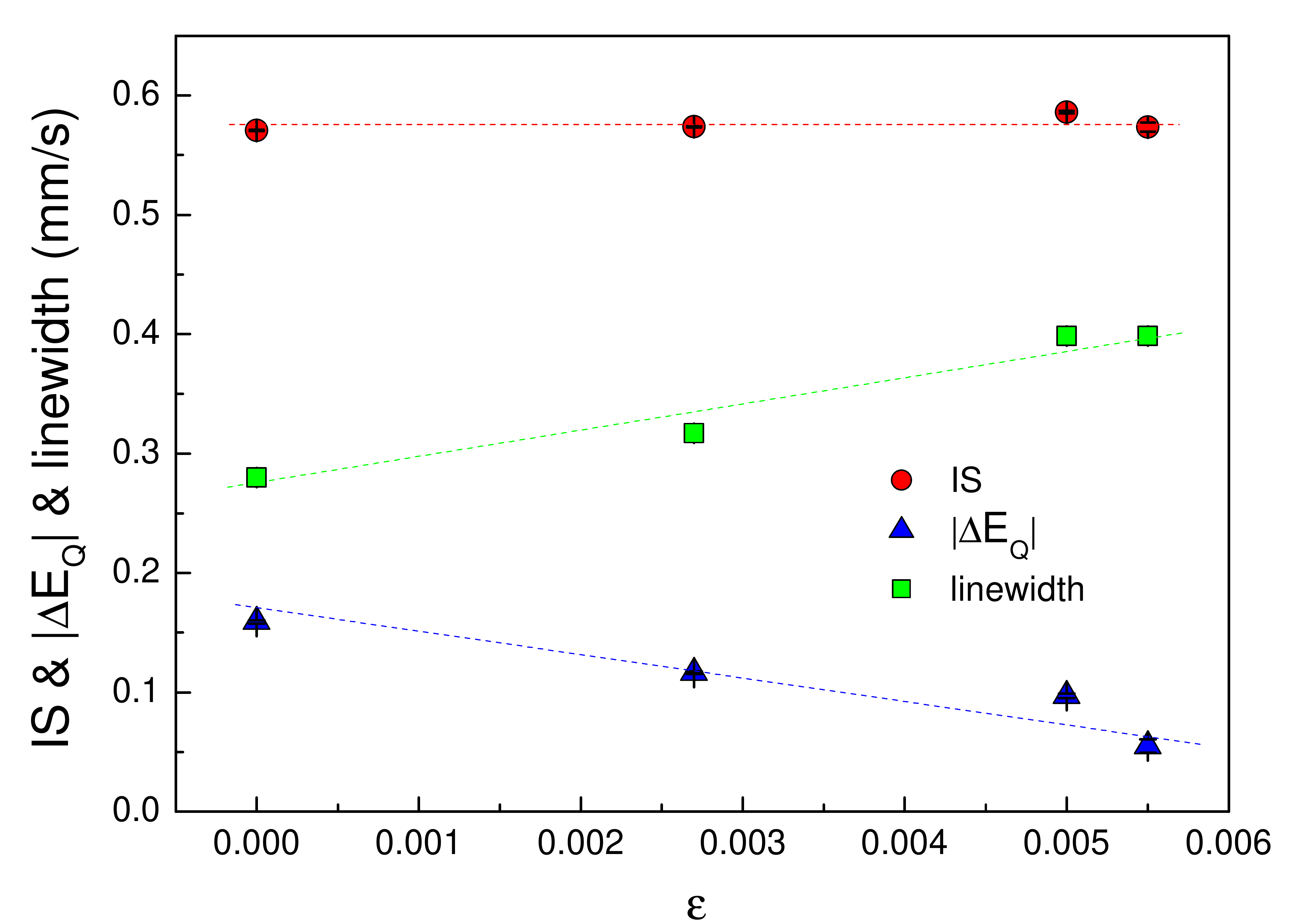}
\caption{\label{fig:epsart} (Color online) the IS, $|$$\Delta$E$_{\mathrm{Q}}$$|$ and linewidth value of the sextet component of the spectra of all four samples at 4.6 K as a function of $\varepsilon$ obtained by fitting XRD result by Eq.(1). Here, the $\varepsilon$ of Sngl. Crys. is assumed as 0. The dashed lines are given as the guides for the eye.}
\end{figure}

The temperature dependent IS, quadrupole splitting ($\Delta$E$_{\mathrm{Q}}$) and linewidth of the sextet of all four samples are summarized in Fig. 8 (a), (b) and (c), respectively. Correspondingly, Fig. 9 is the IS, $|$$\Delta$E$_{\mathrm{Q}}$$|$ and linewidth value of the four samples at 4.6 K as a function of the $\varepsilon$ obtained from the XRD result. Usually, linewidth is able to scale the disorder of environment of $^{57}$Fe nucleus. As expected, the linewidth at 4.6 K increases with the increasing strain. In Fig. 8 (a), the values of IS of all four sample are similar and are barely affected by the strain in the whole temperature range, which indicates the strain in the magnetic phase does not change the electronic density at $^{57}$Fe nucleus position in a significant manner. However, as shown in Fig. 8 (b) and Fig. 9, the $\Delta$E$_{\mathrm{Q}}$ is highly dependent on the degree of the strain. The absolute value of $\Delta$E$_{\mathrm{Q}}$ linearly decrease with increasing strain, that suggests that  strain tends to make the Fe atoms environment more symmetric.

\begin{figure}[htp]
\includegraphics[width=0.5\textwidth]{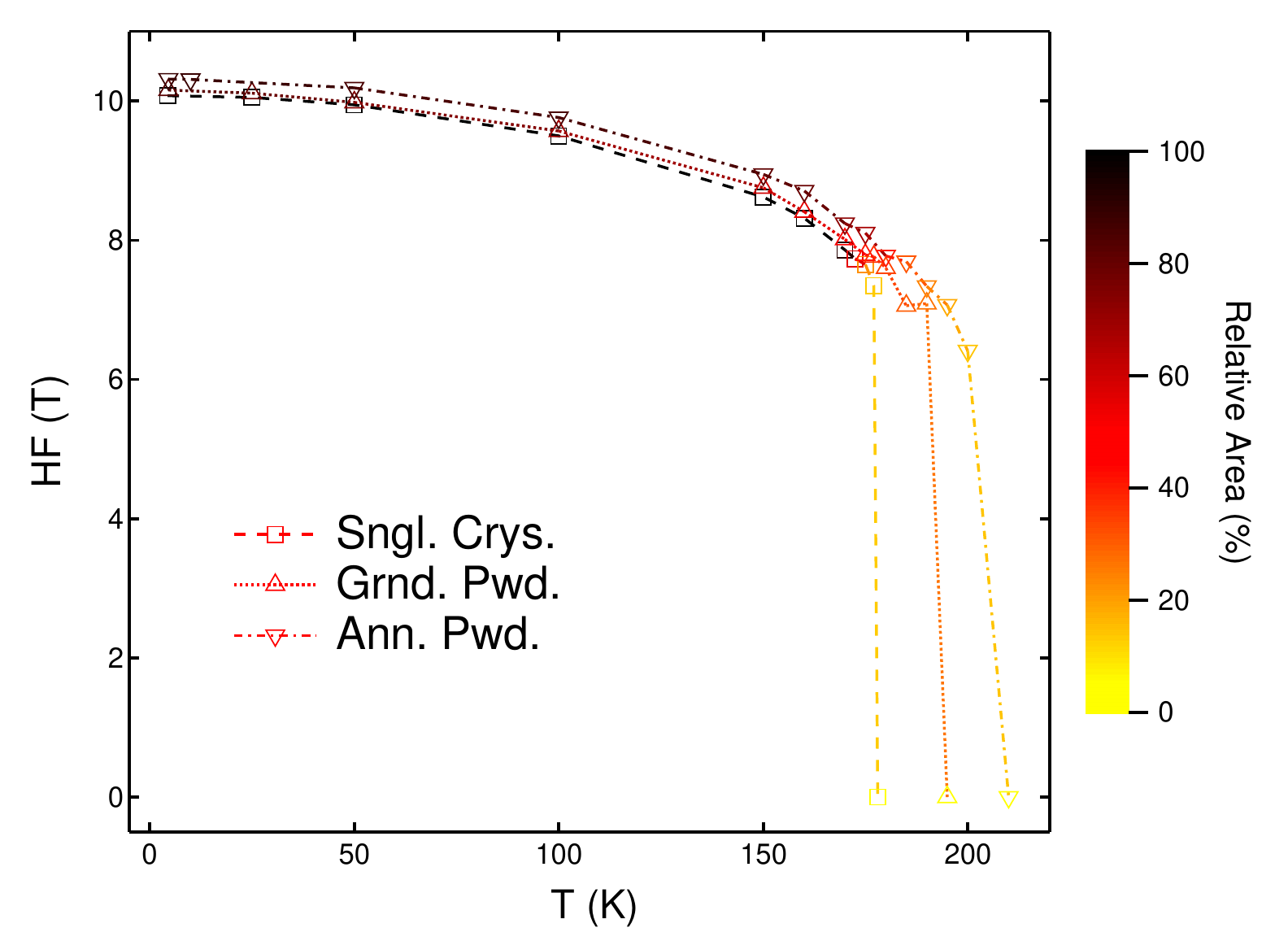}
\caption{\label{fig:epsart} (Color online) Temperature dependent HF of Sngl. Crys., Grnd. Pwd.*, Grnd. Pwd. and Ann. Pwd. The color indicates the relative area of the sextet sub-spectrum as scaled by the right side color column. }
\end{figure}

\begin{figure}[htp]
\includegraphics[width=0.5\textwidth]{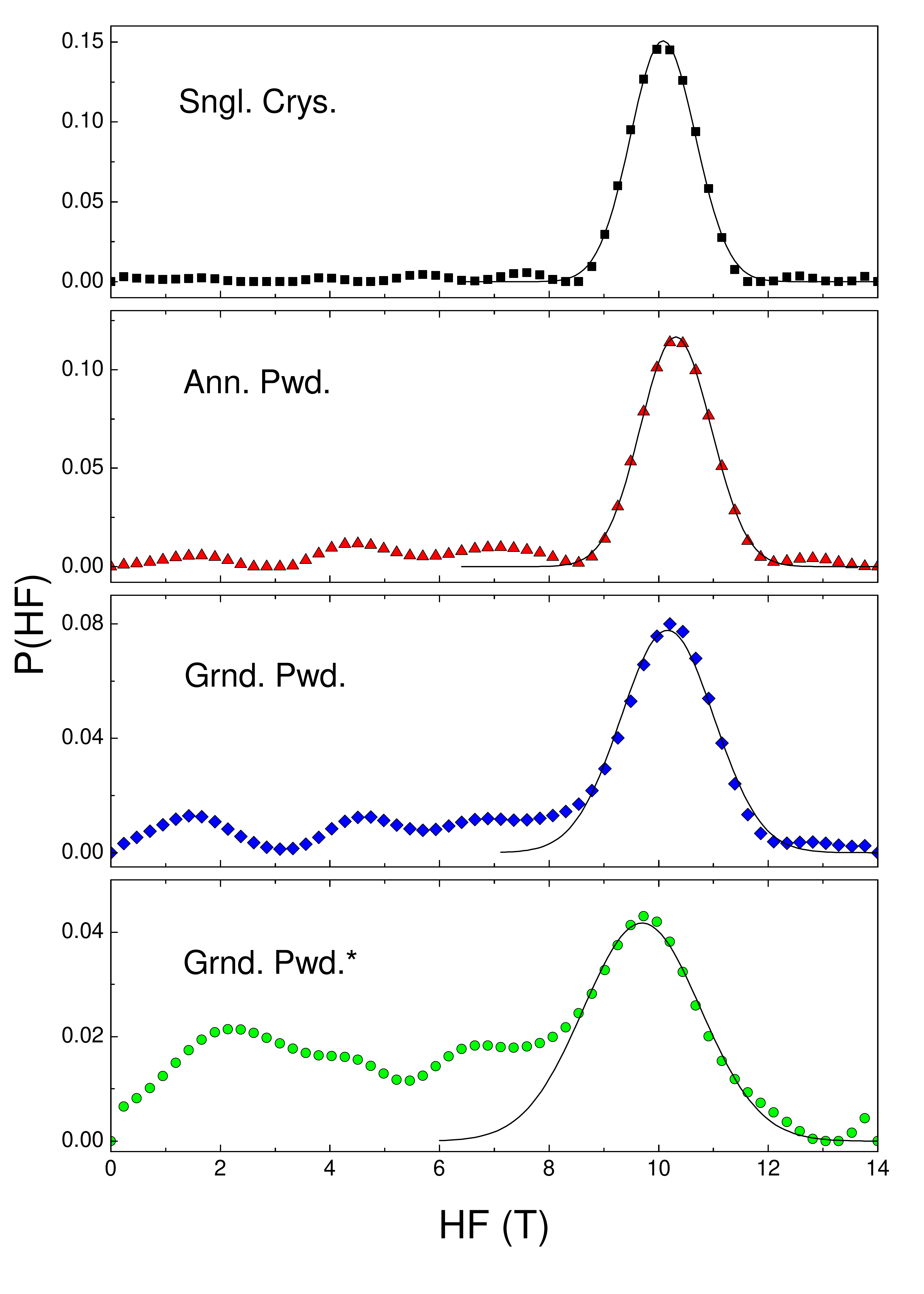}
\caption{\label{fig:epsart} (Color online) Field distribution curves of Sngl. Crys., Grnd. Pwd.*, Grnd. Pwd. and Ann. Pwd. by Hesse-R\"ubartsch method. The solid lines are the results of fitting the main peaks by Gaussian model.}
\end{figure}

The hyperfine field (HF) versus temperature results of all samples are shown in Fig. 10, and the color indicate the relative area of the sextet sub-spectrum. For Sngl. Crys., the HF is $\sim$ 10.1 T at 4.6 K corresponding to $\sim$ 0.78 $\mu_B$ moment on Fe, under the assumption of 13 T/$\mu_B$, which is close to previous reported results \cite{CaMoss1, CaMoss2}. The temperature dependent HF curves of Grnd. Pwd.*, Grnd. Pwd. and Ann. Pwd., are very close to single crystal one. This suggests the magnetic moment associated with Fe in the spin density wave (SDW) state is very robust and is barely affected by the strain.

Based on the temperature and pressure phase diagram of CaFe$_2$As$_2$ \cite{Yu}, for the hydrostatic pressures over 0.4 GPa, the magnetic/structural phase transition will be totally suppressed. As a result of grinding (or grinding and annealing) doublet subspectra are still present at at 4.6 K for Grnd. Pwd.*, Grnd. Pwd., and Ann. Pwd., so in all three samples there are grains under stress high enough to suppress magnetism. On the other hand, in the Grnd. Pwd. and Ann. Pwd. samples a sextet subspectra are still resolvable at temperatures above the transition temperature in CaFe$_2$As$_2$ single crystals (Figs. 6, 7). This means that complex stresses can enhance the magnetic transition temperature and a small, but detectable number of such grains with such stresses are present in these two samples.

In the analysis above, the distribution of strains/properties in and between the grains of the same sample was accounted for by the linewidth. In a different approach,  we can consider a distribution of the hyperfine field caused by strain. We use this approach for the 4.6 K data for all four samples. The Hesse-R\"ubartsch method was  employed to resolve the field distribution of the samples at 4.6 K \cite{Hesse}. Sixty elementary sextets with the same value of IS and $\Delta$E$_{\mathrm{Q}}$ were used. The HF was restricted to less than 14 T and only positive values were considered, with distributions that approach zero at the 14 T. To avoid overfitting, the smoothing factor was fixed at 5, which is usually considered as a moderate value. The obtained field distribution results of all four samples at 4.6 K are shown in Fig. 11. As expected, the distribution of Sngl. Crys. presents a peak around 10 T. Due to existence of the paramagnetic phase for Ann. Pwd, Grnd. Pwd. and Grnd. Pwd.* at 4.6 K, besides the peak around 10 T, there is some fluctuation in low field range as well. For Grnd. Pwd* sample,  a significant contribution of the paramagnetic phase (doublet in the spectrum), accounts for the fluctuation in the low field range. In order to obtain more precise information about the magnetic phase, the main peaks around 10 T were fitted by a Gaussian line for all the samples. The obtained center positions of the peaks are 10.08(1), 10.31(3), 10.16(3) and 9.70(3) and the FWHM are 1.39 T, 1.52 T, 2.00 T and 2.52 T for Sngl. Crys., Ann. Pwd., Grnd. Pwd. and Grnd. Pwd.*, respectively. The center  positions are consistent with the HF results obtained by fitting the spectra with unique field value (shown in Table 2). The evolution of the calculated FWHM of the main peak in the HF distribution also suggests that the strain can result in a distribution of HF, but this is a relatively smaller effect than the change in doublet/sextet relative area shown in Fig. 7. These results suggest that the fits with a single value of the hyperfine field capture the main results of grinding. 

\section{Conclusions}
In summary, $^{57}$Fe M\"ossbauer spectra of CaFe$_2$As$_2$ single crystal, two ground single crystal powder samples, and the annealed ground single crystal powder were collected from 4.6 K to RT. M\"ossbauer spectra can trace the temperature dependent ratio of the magnetic and paramagnetic phases. As a result of grinding, the  paramagnetic phase persists to  low temperatures and its relative amount increases with the increasing strain.  The complex disorder induced by grinding also is able to cause a slight increase of the magnetic phase transition temperature, that is not disappear after annealing. Additionally, the strain does not influence the absolute value of the hyperfine magnetic field (e.g. the magnetic moment on Fe). Once the long range SDW order forms in the compound, the magnetic moment on Fe is  robust, around 0.78 $\mu_B$ at 4.6 K. The IS of the magnetic phase in all the four samples are similar, which indicates the electronic density at the $^{57}$Fe nucleus position is not affected by the strain, but the environment of Fe atom (as reflected in  $|$$\Delta$E$_{\mathrm{Q}}$$|$)  becomes more symmetric with increasing strain.

\begin{acknowledgments}
The authors thank V. Taufour for help with figures and Prof. Hua Pang for useful discussion. X.M. was supported in part by the China Scholarship Council. Work at the Ames Laboratory was supported by the US Department of Energy, Basic Energy Sciences, Division of Materials Sciences and Engineering under Contract no. DE-AC02-07CH11358.
\end{acknowledgments}
\nocite{*}

\providecommand{\noopsort}[1]{}\providecommand{\singleletter}[1]{#1}%

\begin{thebibliography}{21}%
\makeatletter
\providecommand \@ifxundefined [1]{%
 \@ifx{#1\undefined}
}%
\providecommand \@ifnum [1]{%
 \ifnum #1\expandafter \@firstoftwo
 \else \expandafter \@secondoftwo
 \fi
}%
\providecommand \@ifx [1]{%
 \ifx #1\expandafter \@firstoftwo
 \else \expandafter \@secondoftwo
 \fi
}%
\providecommand \natexlab [1]{#1}%
\providecommand \enquote  [1]{``#1''}%
\providecommand \bibnamefont  [1]{#1}%
\providecommand \bibfnamefont [1]{#1}%
\providecommand \citenamefont [1]{#1}%
\providecommand \href@noop [0]{\@secondoftwo}%
\providecommand \href [0]{\begingroup \@sanitize@url \@href}%
\providecommand \@href[1]{\@@startlink{#1}\@@href}%
\providecommand \@@href[1]{\endgroup#1\@@endlink}%
\providecommand \@sanitize@url [0]{\catcode `\\12\catcode `\$12\catcode
  `\&12\catcode `\#12\catcode `\^12\catcode `\_12\catcode `\%12\relax}%
\providecommand \@@startlink[1]{}%
\providecommand \@@endlink[0]{}%
\providecommand \url  [0]{\begingroup\@sanitize@url \@url }%
\providecommand \@url [1]{\endgroup\@href {#1}{\urlprefix }}%
\providecommand \urlprefix  [0]{URL }%
\providecommand \Eprint [0]{\href }%
\providecommand \doibase [0]{http://dx.doi.org/}%
\providecommand \selectlanguage [0]{\@gobble}%
\providecommand \bibinfo  [0]{\@secondoftwo}%
\providecommand \bibfield  [0]{\@secondoftwo}%
\providecommand \translation [1]{[#1]}%
\providecommand \BibitemOpen [0]{}%
\providecommand \bibitemStop [0]{}%
\providecommand \bibitemNoStop [0]{.\EOS\space}%
\providecommand \EOS [0]{\spacefactor3000\relax}%
\providecommand \BibitemShut  [1]{\csname bibitem#1\endcsname}%
\let\auto@bib@innerbib\@empty
%</preamble>
\bibitem [{\citenamefont {Ni}\ \emph {et~al.}(2008)\citenamefont {Ni},
  \citenamefont {Nandi}, \citenamefont {Kreyssig}, \citenamefont {Goldman},
  \citenamefont {Mun}, \citenamefont {Bud'ko},\ and\ \citenamefont
  {Canfield}}]{NiNiFirst}%
  \BibitemOpen
  \bibfield  {author} {\bibinfo {author} {\bibfnamefont {N.}~\bibnamefont
  {Ni}}, \bibinfo {author} {\bibfnamefont {S.}~\bibnamefont {Nandi}}, \bibinfo
  {author} {\bibfnamefont {A.}~\bibnamefont {Kreyssig}}, \bibinfo {author}
  {\bibfnamefont {A.~I.}\ \bibnamefont {Goldman}}, \bibinfo {author}
  {\bibfnamefont {E.~D.}\ \bibnamefont {Mun}}, \bibinfo {author} {\bibfnamefont
  {S.~L.}\ \bibnamefont {Bud'ko}}, \ and\ \bibinfo {author} {\bibfnamefont
  {P.~C.}\ \bibnamefont {Canfield}},\ }\href@noop {} {\bibfield  {journal}
  {\bibinfo  {journal} {Phys. Rev. B}\ }\textbf {\bibinfo {volume} {78}},\
  \bibinfo {pages} {014523} (\bibinfo {year} {2008})}\BibitemShut {NoStop}%
\bibitem [{\citenamefont {Goldman}\ \emph {et~al.}(2008)\citenamefont
  {Goldman}, \citenamefont {Argyriou}, \citenamefont {Ouladdiaf}, \citenamefont
  {Chatterji}, \citenamefont {Kreyssig}, \citenamefont {Nandi}, \citenamefont
  {Ni}, \citenamefont {Bud'ko}, \citenamefont {Canfield},\ and\ \citenamefont
  {McQueeney}}]{Coupling}%
  \BibitemOpen
  \bibfield  {author} {\bibinfo {author} {\bibfnamefont {A.~I.}\ \bibnamefont
  {Goldman}}, \bibinfo {author} {\bibfnamefont {D.~N.}\ \bibnamefont
  {Argyriou}}, \bibinfo {author} {\bibfnamefont {B.}~\bibnamefont {Ouladdiaf}},
  \bibinfo {author} {\bibfnamefont {T.}~\bibnamefont {Chatterji}}, \bibinfo
  {author} {\bibfnamefont {A.}~\bibnamefont {Kreyssig}}, \bibinfo {author}
  {\bibfnamefont {S.}~\bibnamefont {Nandi}}, \bibinfo {author} {\bibfnamefont
  {N.}~\bibnamefont {Ni}}, \bibinfo {author} {\bibfnamefont {S.~L.}\
  \bibnamefont {Bud'ko}}, \bibinfo {author} {\bibfnamefont {P.~C.}\
  \bibnamefont {Canfield}}, \ and\ \bibinfo {author} {\bibfnamefont {R.~J.}\
  \bibnamefont {McQueeney}},\ }\href@noop {} {\bibfield  {journal} {\bibinfo
  {journal} {Phys. Rev. B}\ }\textbf {\bibinfo {volume} {78}},\ \bibinfo
  {pages} {100506} (\bibinfo {year} {2008})}\BibitemShut {NoStop}%
\bibitem [{\citenamefont {Torikachvili}\ \emph {et~al.}(2008)\citenamefont
  {Torikachvili}, \citenamefont {Bud'ko}, \citenamefont {Ni},\ and\
  \citenamefont {Canfield}}]{Milton1}%
  \BibitemOpen
  \bibfield  {author} {\bibinfo {author} {\bibfnamefont {M.~S.}\ \bibnamefont
  {Torikachvili}}, \bibinfo {author} {\bibfnamefont {S.~L.}\ \bibnamefont
  {Bud'ko}}, \bibinfo {author} {\bibfnamefont {N.}~\bibnamefont {Ni}}, \ and\
  \bibinfo {author} {\bibfnamefont {P.~C.}\ \bibnamefont {Canfield}},\
  }\href@noop {} {\bibfield  {journal} {\bibinfo  {journal} {Phys. Rev. Lett.}\
  }\textbf {\bibinfo {volume} {101}},\ \bibinfo {pages} {057006} (\bibinfo
  {year} {2008})}\BibitemShut {NoStop}%
\bibitem [{\citenamefont {Yu}\ \emph {et~al.}(2009)\citenamefont {Yu},
  \citenamefont {Aczel}, \citenamefont {Williams}, \citenamefont {Bud'ko},
  \citenamefont {Ni}, \citenamefont {Canfield},\ and\ \citenamefont
  {Luke}}]{Yu}%
  \BibitemOpen
  \bibfield  {author} {\bibinfo {author} {\bibfnamefont {W.}~\bibnamefont
  {Yu}}, \bibinfo {author} {\bibfnamefont {A.~A.}\ \bibnamefont {Aczel}},
  \bibinfo {author} {\bibfnamefont {T.~J.}\ \bibnamefont {Williams}}, \bibinfo
  {author} {\bibfnamefont {S.~L.}\ \bibnamefont {Bud'ko}}, \bibinfo {author}
  {\bibfnamefont {N.}~\bibnamefont {Ni}}, \bibinfo {author} {\bibfnamefont
  {P.~C.}\ \bibnamefont {Canfield}}, \ and\ \bibinfo {author} {\bibfnamefont
  {G.~M.}\ \bibnamefont {Luke}},\ }\href@noop {} {\bibfield  {journal}
  {\bibinfo  {journal} {Phys. Rev. B}\ }\textbf {\bibinfo {volume} {79}},\
  \bibinfo {pages} {020511} (\bibinfo {year} {2009})}\BibitemShut {NoStop}%
\bibitem [{\citenamefont {Kreyssig}\ \emph {et~al.}(2008)\citenamefont
  {Kreyssig}, \citenamefont {Green}, \citenamefont {Lee}, \citenamefont
  {Samolyuk}, \citenamefont {Zajdel}, \citenamefont {Lynn}, \citenamefont
  {Bud'ko}, \citenamefont {Torikachvili}, \citenamefont {Ni}, \citenamefont
  {Nandi}, \citenamefont {Le{\~{a}}o}, \citenamefont {Poulton}, \citenamefont
  {Argyriou}, \citenamefont {Harmon}, \citenamefont {McQueeney}, \citenamefont
  {Canfield},\ and\ \citenamefont {Goldman}}]{scattering}%
  \BibitemOpen
  \bibfield  {author} {\bibinfo {author} {\bibfnamefont {A.}~\bibnamefont
  {Kreyssig}}, \bibinfo {author} {\bibfnamefont {M.~A.}\ \bibnamefont {Green}},
  \bibinfo {author} {\bibfnamefont {Y.}~\bibnamefont {Lee}}, \bibinfo {author}
  {\bibfnamefont {G.~D.}\ \bibnamefont {Samolyuk}}, \bibinfo {author}
  {\bibfnamefont {P.}~\bibnamefont {Zajdel}}, \bibinfo {author} {\bibfnamefont
  {J.~W.}\ \bibnamefont {Lynn}}, \bibinfo {author} {\bibfnamefont {S.~L.}\
  \bibnamefont {Bud'ko}}, \bibinfo {author} {\bibfnamefont {M.~S.}\
  \bibnamefont {Torikachvili}}, \bibinfo {author} {\bibfnamefont
  {N.}~\bibnamefont {Ni}}, \bibinfo {author} {\bibfnamefont {S.}~\bibnamefont
  {Nandi}}, \bibinfo {author} {\bibfnamefont {J.~B.}\ \bibnamefont
  {Le{\~{a}}o}}, \bibinfo {author} {\bibfnamefont {S.~J.}\ \bibnamefont
  {Poulton}}, \bibinfo {author} {\bibfnamefont {D.~N.}\ \bibnamefont
  {Argyriou}}, \bibinfo {author} {\bibfnamefont {B.~N.}\ \bibnamefont
  {Harmon}}, \bibinfo {author} {\bibfnamefont {R.~J.}\ \bibnamefont
  {McQueeney}}, \bibinfo {author} {\bibfnamefont {P.~C.}\ \bibnamefont
  {Canfield}}, \ and\ \bibinfo {author} {\bibfnamefont {A.~I.}\ \bibnamefont
  {Goldman}},\ }\href@noop {} {\bibfield  {journal} {\bibinfo  {journal} {Phys.
  Rev. B}\ }\textbf {\bibinfo {volume} {78}},\ \bibinfo {pages} {184517}
  (\bibinfo {year} {2008})}\BibitemShut {NoStop}%
\bibitem [{\citenamefont {Torikachvili}\ \emph {et~al.}(2009)\citenamefont
  {Torikachvili}, \citenamefont {Bud'ko}, \citenamefont {Ni}, \citenamefont
  {Canfield},\ and\ \citenamefont {Hannahs}}]{Milton2}%
  \BibitemOpen
  \bibfield  {author} {\bibinfo {author} {\bibfnamefont {M.~S.}\ \bibnamefont
  {Torikachvili}}, \bibinfo {author} {\bibfnamefont {S.~L.}\ \bibnamefont
  {Bud'ko}}, \bibinfo {author} {\bibfnamefont {N.}~\bibnamefont {Ni}}, \bibinfo
  {author} {\bibfnamefont {P.~C.}\ \bibnamefont {Canfield}}, \ and\ \bibinfo
  {author} {\bibfnamefont {S.~T.}\ \bibnamefont {Hannahs}},\ }\href@noop {}
  {\bibfield  {journal} {\bibinfo  {journal} {Phys. Rev. B}\ }\textbf {\bibinfo
  {volume} {80}},\ \bibinfo {pages} {014521} (\bibinfo {year}
  {2009})}\BibitemShut {NoStop}%
\bibitem [{\citenamefont {Baek}\ \emph {et~al.}(2009)\citenamefont {Baek},
  \citenamefont {Lee}, \citenamefont {Brown}, \citenamefont {Curro},
  \citenamefont {Bauer}, \citenamefont {Ronning}, \citenamefont {Park},\ and\
  \citenamefont {Thompson}}]{LANL}%
  \BibitemOpen
  \bibfield  {author} {\bibinfo {author} {\bibfnamefont {S.-H.}\ \bibnamefont
  {Baek}}, \bibinfo {author} {\bibfnamefont {H.}~\bibnamefont {Lee}}, \bibinfo
  {author} {\bibfnamefont {S.~E.}\ \bibnamefont {Brown}}, \bibinfo {author}
  {\bibfnamefont {N.~J.}\ \bibnamefont {Curro}}, \bibinfo {author}
  {\bibfnamefont {E.~D.}\ \bibnamefont {Bauer}}, \bibinfo {author}
  {\bibfnamefont {F.}~\bibnamefont {Ronning}}, \bibinfo {author} {\bibfnamefont
  {T.}~\bibnamefont {Park}}, \ and\ \bibinfo {author} {\bibfnamefont {J.~D.}\
  \bibnamefont {Thompson}},\ }\href@noop {} {\bibfield  {journal} {\bibinfo
  {journal} {Phys. Rev. Lett.}\ }\textbf {\bibinfo {volume} {102}},\ \bibinfo
  {pages} {227601} (\bibinfo {year} {2009})}\BibitemShut {NoStop}%
\bibitem [{\citenamefont {Canfield}\ \emph {et~al.}(2009)\citenamefont
  {Canfield}, \citenamefont {Bud'ko}, \citenamefont {Ni}, \citenamefont
  {Kreyssig}, \citenamefont {Goldman}, \citenamefont {McQueeney}, \citenamefont
  {Torikachvili}, \citenamefont {Argyriou}, \citenamefont {Luke},\ and\
  \citenamefont {Yu}}]{PaulPhysicaC}%
  \BibitemOpen
  \bibfield  {author} {\bibinfo {author} {\bibfnamefont {P.}~\bibnamefont
  {Canfield}}, \bibinfo {author} {\bibfnamefont {S.}~\bibnamefont {Bud'ko}},
  \bibinfo {author} {\bibfnamefont {N.}~\bibnamefont {Ni}}, \bibinfo {author}
  {\bibfnamefont {A.}~\bibnamefont {Kreyssig}}, \bibinfo {author}
  {\bibfnamefont {A.}~\bibnamefont {Goldman}}, \bibinfo {author} {\bibfnamefont
  {R.}~\bibnamefont {McQueeney}}, \bibinfo {author} {\bibfnamefont
  {M.}~\bibnamefont {Torikachvili}}, \bibinfo {author} {\bibfnamefont
  {D.}~\bibnamefont {Argyriou}}, \bibinfo {author} {\bibfnamefont
  {G.}~\bibnamefont {Luke}}, \ and\ \bibinfo {author} {\bibfnamefont
  {W.}~\bibnamefont {Yu}},\ }\href@noop {} {\bibfield  {journal} {\bibinfo
  {journal} {Physica C}\ }\textbf {\bibinfo {volume} {469}},\ \bibinfo {pages}
  {404} (\bibinfo {year} {2009})}\BibitemShut {NoStop}%
\bibitem [{\citenamefont {Ran}\ \emph {et~al.}(2012)\citenamefont {Ran},
  \citenamefont {Bud'ko}, \citenamefont {Straszheim}, \citenamefont {Soh},
  \citenamefont {Kim}, \citenamefont {Kreyssig}, \citenamefont {Goldman},\ and\
  \citenamefont {Canfield}}]{Sheng2}%
  \BibitemOpen
  \bibfield  {author} {\bibinfo {author} {\bibfnamefont {S.}~\bibnamefont
  {Ran}}, \bibinfo {author} {\bibfnamefont {S.~L.}\ \bibnamefont {Bud'ko}},
  \bibinfo {author} {\bibfnamefont {W.~E.}\ \bibnamefont {Straszheim}},
  \bibinfo {author} {\bibfnamefont {J.}~\bibnamefont {Soh}}, \bibinfo {author}
  {\bibfnamefont {M.~G.}\ \bibnamefont {Kim}}, \bibinfo {author} {\bibfnamefont
  {A.}~\bibnamefont {Kreyssig}}, \bibinfo {author} {\bibfnamefont {A.~I.}\
  \bibnamefont {Goldman}}, \ and\ \bibinfo {author} {\bibfnamefont {P.~C.}\
  \bibnamefont {Canfield}},\ }\href@noop {} {\bibfield  {journal} {\bibinfo
  {journal} {Phys. Rev. B}\ }\textbf {\bibinfo {volume} {85}},\ \bibinfo
  {pages} {224528} (\bibinfo {year} {2012})}\BibitemShut {NoStop}%
\bibitem [{\citenamefont {Bud'ko}\ \emph {et~al.}(2013)\citenamefont {Bud'ko},
  \citenamefont {Ran},\ and\ \citenamefont {Canfield}}]{TE}%
  \BibitemOpen
  \bibfield  {author} {\bibinfo {author} {\bibfnamefont {S.~L.}\ \bibnamefont
  {Bud'ko}}, \bibinfo {author} {\bibfnamefont {S.}~\bibnamefont {Ran}}, \ and\
  \bibinfo {author} {\bibfnamefont {P.~C.}\ \bibnamefont {Canfield}},\
  }\href@noop {} {\bibfield  {journal} {\bibinfo  {journal} {Phys. Rev. B}\
  }\textbf {\bibinfo {volume} {88}},\ \bibinfo {pages} {064513} (\bibinfo
  {year} {2013})}\BibitemShut {NoStop}%
\bibitem [{\citenamefont {Gati}\ \emph {et~al.}(2012)\citenamefont {Gati},
  \citenamefont {K{\"o}hler}, \citenamefont {Guterding}, \citenamefont {Wolf},
  \citenamefont {Kn{\"o}ner}, \citenamefont {Ran}, \citenamefont {Bud'ko},
  \citenamefont {Canfield},\ and\ \citenamefont {Lang}}]{gati}%
  \BibitemOpen
  \bibfield  {author} {\bibinfo {author} {\bibfnamefont {E.}~\bibnamefont
  {Gati}}, \bibinfo {author} {\bibfnamefont {S.}~\bibnamefont {K{\"o}hler}},
  \bibinfo {author} {\bibfnamefont {D.}~\bibnamefont {Guterding}}, \bibinfo
  {author} {\bibfnamefont {B.}~\bibnamefont {Wolf}}, \bibinfo {author}
  {\bibfnamefont {S.}~\bibnamefont {Kn{\"o}ner}}, \bibinfo {author}
  {\bibfnamefont {S.}~\bibnamefont {Ran}}, \bibinfo {author} {\bibfnamefont
  {S.~L.}\ \bibnamefont {Bud'ko}}, \bibinfo {author} {\bibfnamefont {P.~C.}\
  \bibnamefont {Canfield}}, \ and\ \bibinfo {author} {\bibfnamefont
  {M.}~\bibnamefont {Lang}},\ }\href@noop {} {\bibfield  {journal} {\bibinfo
  {journal} {Phys. Rev. B}\ }\textbf {\bibinfo {volume} {86}},\ \bibinfo
  {pages} {220511} (\bibinfo {year} {2012})}\BibitemShut {NoStop}%
\bibitem [{\citenamefont {Ran}\ \emph {et~al.}(2011)\citenamefont {Ran},
  \citenamefont {Bud'ko}, \citenamefont {Pratt}, \citenamefont {Kreyssig},
  \citenamefont {Kim}, \citenamefont {Kramer}, \citenamefont {Ryan},
  \citenamefont {Rowan-Weetaluktuk}, \citenamefont {Furukawa}, \citenamefont
  {Roy}, \citenamefont {Goldman},\ and\ \citenamefont {Canfield}}]{Sheng1}%
  \BibitemOpen
  \bibfield  {author} {\bibinfo {author} {\bibfnamefont {S.}~\bibnamefont
  {Ran}}, \bibinfo {author} {\bibfnamefont {S.~L.}\ \bibnamefont {Bud'ko}},
  \bibinfo {author} {\bibfnamefont {D.~K.}\ \bibnamefont {Pratt}}, \bibinfo
  {author} {\bibfnamefont {A.}~\bibnamefont {Kreyssig}}, \bibinfo {author}
  {\bibfnamefont {M.~G.}\ \bibnamefont {Kim}}, \bibinfo {author} {\bibfnamefont
  {M.~J.}\ \bibnamefont {Kramer}}, \bibinfo {author} {\bibfnamefont {D.~H.}\
  \bibnamefont {Ryan}}, \bibinfo {author} {\bibfnamefont {W.~N.}\ \bibnamefont
  {Rowan-Weetaluktuk}}, \bibinfo {author} {\bibfnamefont {Y.}~\bibnamefont
  {Furukawa}}, \bibinfo {author} {\bibfnamefont {B.}~\bibnamefont {Roy}},
  \bibinfo {author} {\bibfnamefont {A.~I.}\ \bibnamefont {Goldman}}, \ and\
  \bibinfo {author} {\bibfnamefont {P.~C.}\ \bibnamefont {Canfield}},\
  }\href@noop {} {\bibfield  {journal} {\bibinfo  {journal} {Phys. Rev. B}\
  }\textbf {\bibinfo {volume} {83}},\ \bibinfo {pages} {144517} (\bibinfo
  {year} {2011})}\BibitemShut {NoStop}%
\bibitem [{\citenamefont {Ran}(2014)}]{ShengThesis}%
  \BibitemOpen
  \bibfield  {author} {\bibinfo {author} {\bibfnamefont {S.}~\bibnamefont
  {Ran}},\ }\href@noop {} {\bibfield  {journal} {\bibinfo  {journal} {Combined
  effects of post-growth thermal treatment and chemical substitution on
  physical properties of CaFe$_2$As$_2$, Ph.D. Thesis, Iowa State University}\
  } (\bibinfo {year} {2014})}\BibitemShut {NoStop}%
\bibitem [{\citenamefont {Canfield}\ and\ \citenamefont
  {Fisk}(1992)}]{CanfieldFisk}%
  \BibitemOpen
  \bibfield  {author} {\bibinfo {author} {\bibfnamefont {P.}~\bibnamefont
  {Canfield}}\ and\ \bibinfo {author} {\bibfnamefont {Z.}~\bibnamefont
  {Fisk}},\ }\href@noop {} {\bibfield  {journal} {\bibinfo  {journal} {Philos.
  Mag. B}\ }\textbf {\bibinfo {volume} {65}},\ \bibinfo {pages} {1117}
  (\bibinfo {year} {1992})}\BibitemShut {NoStop}%
\bibitem [{\citenamefont {Toby}(2001)}]{EXPGUI}%
  \BibitemOpen
  \bibfield  {author} {\bibinfo {author} {\bibfnamefont {B.~H.}\ \bibnamefont
  {Toby}},\ }\href@noop {} {\bibfield  {journal} {\bibinfo  {journal} {J. Appl.
  Cryst.}\ }\textbf {\bibinfo {volume} {34}},\ \bibinfo {pages} {210} (\bibinfo
  {year} {2001})}\BibitemShut {NoStop}%
\bibitem [{\citenamefont {Scherrer}()}]{Scherrer}%
  \BibitemOpen
  \bibfield  {author} {\bibinfo {author} {\bibfnamefont {P.}~\bibnamefont
  {Scherrer}},\ }\href@noop {} {\bibinfo  {journal} {\emph{G{\"o}ttinger
  Nachrichten} (1918) R. Zsigmondy, Kolloidchemie (3rd Ed. 1920)}\ ,\ \bibinfo
  {pages} {394}}\BibitemShut {NoStop}%
\bibitem [{\citenamefont {Stokes}\ and\ \citenamefont {Wilson}(1943)}]{Stokes}%
  \BibitemOpen
\bibfield  {journal} {  }\bibfield  {author} {\bibinfo {author} {\bibfnamefont
  {A.~R.}\ \bibnamefont {Stokes}}\ and\ \bibinfo {author} {\bibfnamefont
  {A.~J.~C.}\ \bibnamefont {Wilson}},\ }\href@noop {} {\bibfield  {journal}
  {\bibinfo  {journal} {Proc. Phys. Soc.}\ }\textbf {\bibinfo {volume} {56}},\
  \bibinfo {pages} {174} (\bibinfo {year} {1943})}\BibitemShut {NoStop}%
\bibitem [{\citenamefont {Karen}\ and\ \citenamefont
  {Woodward}(1998)}]{equation1}%
  \BibitemOpen
  \bibfield  {author} {\bibinfo {author} {\bibfnamefont {P.}~\bibnamefont
  {Karen}}\ and\ \bibinfo {author} {\bibfnamefont {P.~M.}\ \bibnamefont
  {Woodward}},\ }\href@noop {} {\bibfield  {journal} {\bibinfo  {journal} {J.
  Solid State Chem.}\ }\textbf {\bibinfo {volume} {141}},\ \bibinfo {pages}
  {78} (\bibinfo {year} {1998})}\BibitemShut {NoStop}%
\bibitem [{\citenamefont {Alzamora}\ \emph {et~al.}(2011)\citenamefont
  {Alzamora}, \citenamefont {Munevar}, \citenamefont {Baggio-Saitovitch},
  \citenamefont {Bud'ko}, \citenamefont {Ni}, \citenamefont {Canfield},\ and\
  \citenamefont {S{\'{a}}nchez}}]{CaMoss1}%
  \BibitemOpen
  \bibfield  {author} {\bibinfo {author} {\bibfnamefont {M.}~\bibnamefont
  {Alzamora}}, \bibinfo {author} {\bibfnamefont {J.}~\bibnamefont {Munevar}},
  \bibinfo {author} {\bibfnamefont {E.}~\bibnamefont {Baggio-Saitovitch}},
  \bibinfo {author} {\bibfnamefont {S.~L.}\ \bibnamefont {Bud'ko}}, \bibinfo
  {author} {\bibfnamefont {N.}~\bibnamefont {Ni}}, \bibinfo {author}
  {\bibfnamefont {P.~C.}\ \bibnamefont {Canfield}}, \ and\ \bibinfo {author}
  {\bibfnamefont {D.~R.}\ \bibnamefont {S{\'{a}}nchez}},\ }\href@noop {}
  {\bibfield  {journal} {\bibinfo  {journal} {J. Phys.: Condens. Matter}\
  }\textbf {\bibinfo {volume} {23}},\ \bibinfo {pages} {145701} (\bibinfo
  {year} {2011})}\BibitemShut {NoStop}%
\bibitem [{\citenamefont {Li}\ \emph {et~al.}(2011)\citenamefont {Li},
  \citenamefont {Ma}, \citenamefont {Pang},\ and\ \citenamefont
  {Li}}]{CaMoss2}%
  \BibitemOpen
  \bibfield  {author} {\bibinfo {author} {\bibfnamefont {Z.}~\bibnamefont
  {Li}}, \bibinfo {author} {\bibfnamefont {X.}~\bibnamefont {Ma}}, \bibinfo
  {author} {\bibfnamefont {H.}~\bibnamefont {Pang}}, \ and\ \bibinfo {author}
  {\bibfnamefont {F.}~\bibnamefont {Li}},\ }\href@noop {} {\bibfield  {journal}
  {\bibinfo  {journal} {J. Phys.: Condens. Matter}\ }\textbf {\bibinfo {volume}
  {23}},\ \bibinfo {pages} {255701} (\bibinfo {year} {2011})}\BibitemShut
  {NoStop}%
\bibitem [{\citenamefont {Hesse}\ and\ \citenamefont
  {R{\"{u}}bartsch}(1974)}]{Hesse}%
  \BibitemOpen
  \bibfield  {author} {\bibinfo {author} {\bibfnamefont {J.}~\bibnamefont
  {Hesse}}\ and\ \bibinfo {author} {\bibfnamefont {A.}~\bibnamefont
  {R{\"{u}}bartsch}},\ }\href@noop {} {\bibfield  {journal} {\bibinfo
  {journal} {J. Phys. E Sci. Instr.}\ }\textbf {\bibinfo {volume} {7}},\
  \bibinfo {pages} {526} (\bibinfo {year} {1974})}\BibitemShut {NoStop}%
\end{thebibliography}
\end{document}